\documentclass[prb,twocolumn]{revtex4}

\usepackage{amsmath}
\usepackage{amssymb}
\usepackage{graphicx}

\usepackage{lineno,hyperref}
\modulolinenumbers[5]

\begin{document}

\title{Magnon cotunneling through a quantum dot}
\author{\L{}ukasz Karwacki}
\email{karwacki@amu.edu.pl}
\affiliation{Faculty of Physics, Adam Mickiewicz University, ul. Umultowska 85, 61-614 Pozna\'{n},
Poland}

\begin{abstract}
I consider a single-level quantum dot coupled to two reservoirs of spin waves (magnons). Such systems have been studied recently from the point of view of possible coupling between electronic and magnonic spin currents. However, usually weakly coupled systems were investigated. When coupling between the dot and reservoirs is not weak, then higher order processes play a role and have to be included. Here I consider cotunneling of magnons through a spin-occupied quantum dot, which can be understood as a magnon (spin) leakage current in analogy to leakage currents in charge-based electronics. Particular emphasis has been put on investigating the effect of magnetic field and temperature difference between the magnonic reservoirs.
\end{abstract}

\maketitle

\section{Introduction}

Recently one can observe an  increasing interest in magnons as possible information carriers for electronic applications. This interest has lead to the emergence of spin caloritronics and magnon spintronics, aimed at search for the possibility of generating and manipulating magnons via a temperature gradient~\cite{Uchida, Uchida2,Serga,Bauer,Xiao,Jaworski,Flipse,Lenk}. One of the most widely used materials as a source of magnons is yttrium-iron garnet (YIG)~\cite{Serga}. An important advantage of this material  is its very low Gilbert damping factor, and thus a long propagation length of magnons, even at room temperatures~\cite{Cornelissen}. Some magnon-based information processing components, like multiplexers or transistors~\cite{Vogt,Chumak}, have been already realized experimentally. Moreover, further devices like spin Seebeck diodes have been proposed~\cite{Borlenghi,Borlenghi2}. However, the fully magnon-based electronic components are still far from implementation on a larger scale. Currently the most important research direction is focused on hybrid electron and magnon based devices. The important issues that arise in applications and theoretical considerations are the problem of interface between insulating and metallic layers, conversion of spin to charge (and vice versa)  currents and choice of quantization axis for noncollinearly magnetized electrodes~\cite{Ren,Ren2,Kajiwara,Zhang,Chotorlishvili,Shekhter}.

One of convenient systems to study the above mentioned processes is a quantum dot connected to two magnonic reservoirs. This is because quantum dot based systems allow  for a relatively easy control of single particle transport by external gates. Recently, quantum dot based heat engines have been proposed~\cite{Esposito2009,Esposito2010,Strelcyk,Sothmann2010,Sothmann2012,Wohlman,Sanchez2013,Bergenfeldt,Sothmann2015,karwackiPRB}, where the temperature gradient  has been shown to drive electron and magnon currents or phonon-assisted and magnon-assisted electron currents. The considerations, however were usually limited to the lowest-order processes, i.e. sequential transport of magnons though quantum dots. Such an approximation is reasonable when coupling between the dot and magnonic reservoirs is small. For a stronger coupling (or when sequential transport is suppressed), one needs to take into account also higher order processes.
In this paper I consider magnon transport in two distinct regimes: (i) when both sequential and cotunneling processes contribute to the total magnon transport, and (ii) when sequential transport is forbidden, while magnons can flow through the quantum dot due to elastic cotunneling processes. The latter processes  can be considered as magnon leakage. In Section 2 I introduce the model and describe the T-Matrix method used to obtain the magnon currents in both considered situations. Numerical results based on this analysis are presented and discussed in Section 3, while final conclusions are summarized in Section 4.

\section{Theoretical description}

\subsection{Model}

\begin{figure}[!t]
\centering
\includegraphics[width=0.8\columnwidth]{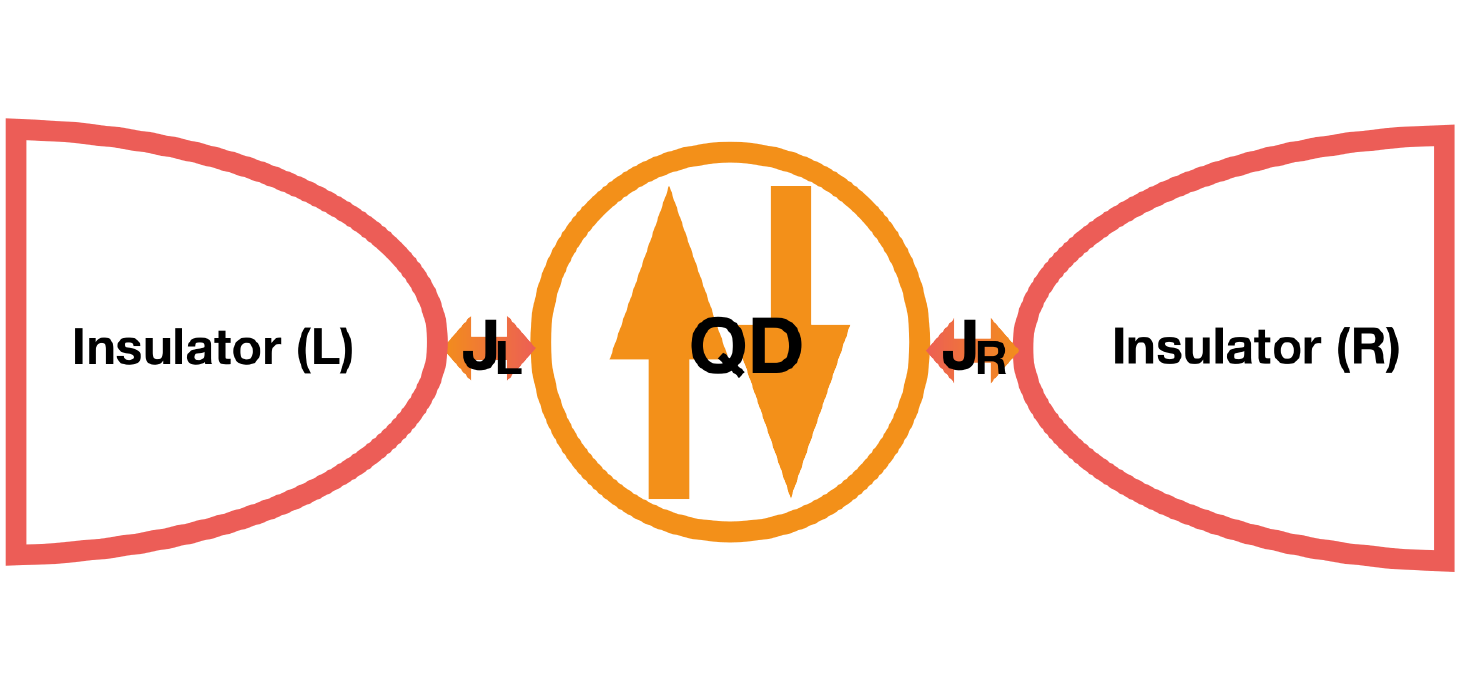}
\caption{Schematic representation of the system considered in this paper. A single-level quantum dot (QD) occupied by either spin $\uparrow$ or spin $\downarrow$ electron is coupled to two magnetic insulators (L,R). The insulating leads are also referred to as magnonic reservoirs. The corresponding coupling strengths are $J_{L}$ and $J_{R}$.}
\label{fig:model}
\end{figure}

The model system studied in this paper is presented schematically in
Fig.~\ref{fig:model}. It is based on a single-level quantum dot which is coupled to two insulating magnetic contacts (reservoirs of magnons). One of the best materials for the magnonic reservoirs might be YIG (yttrium-iron garnet) ferrimagnet due to its very low magnetic damping, and thus long
spin-wave lifetime~\cite{Serga}.

The system can be described by a general Hamiltonian of the form,
\begin{equation}\label{eq:hamiltonian}
H=H_{\rm QD}+H_{\rm mag}+H^{\rm t}_{\rm mag}.
\end{equation}
The first term describes the quantum dot and is assumed in the Anderson form,
\begin{equation}\label{eq:qd}
H_{\rm QD}=\sum_{\sigma}\varepsilon_{d\sigma}d_{\sigma}^{\dagger}d_{\sigma}\,,
\end{equation}
where $\varepsilon_{d\sigma}=\varepsilon_{d}-\hat{\sigma}g\mu_BB/2$ is the
dot's level energy, whose degeneracy is lifted by an external magnetic
field $B$ ($\varepsilon_{d}$ is the bare dot's level energy). Here, $g$ is the Land\'{e} factor for the dot, $\mu_B$ is the Bohr magneton,
while $\hat{\sigma}=+(-)$ for $\sigma=\uparrow(\downarrow)$. Note,  the positive magnetic field $B$ is opposite to the axis $z$.

The magnetic Hamiltonian, 
\begin{equation}
H_{\textrm{mag}}=\frac{1}{2}\sum_{\alpha,i,\delta} J_{\alpha}^{\textrm{ex}}\mathbf{S}_{\alpha ,i}\cdot\mathbf{S}_{\alpha,i+\delta}-g_m\mu_BB\sum_{\alpha,i}S_{\alpha,i}^z,
\end{equation}
describes exchange-coupled localized spins, where $J_\alpha^{\textrm{ex}}$ is the exchange integral and $g_m$ the Land\'{e}'s $g$-factor of magnetic electrodes, and $\mathbf{S}_{\alpha,i}=(S_{\alpha,i}^{x},S_{\alpha,i}^{y},S_{\alpha,i}^{z})$ is the spin of $i$th site in the electrode $\alpha$. Symbol $\delta$ in the sum denotes position of the nearest neighbor of $i$th spin. Moreover, it is assumed that the spins are coupled ferromagnetically, i.e $J_\alpha<0$, and $g_m$ is identical in both electrodes.   

Performing Fourier transform of the above Hamiltonian and introducing the spin ladder operators $S_{\alpha\mathbf{q}}^{\pm}=S_{\alpha\mathbf{q}}^x\pm iS_{\alpha\mathbf{q}}^y$ one can apply Holstein-Primakoff transformation. It consists of expressing spin operators with creation and annihilation operators that obey bosonic commutation rules. The spin operators can be expressed as~\cite{HolsteinPrimakoff}:
\begin{subequations}
\begin{align}
S_{\alpha\mathbf{q}}^{+}&= \hbar\sqrt{2S_\alpha}\sqrt{1-\frac{a_{\alpha\mathbf{q}}^\dagger a_{\alpha\mathbf{q}}}{2S_\alpha}}a_{\alpha\mathbf{q}}\approx \hbar\sqrt{2S_\alpha}a_{\alpha\mathbf{q}}\,, \\  
S_{\alpha\mathbf{q}}^{-}&= \hbar\sqrt{2S_\alpha}a_{\alpha\mathbf{q}}^\dagger\sqrt{1-\frac{a_{\alpha\mathbf{q}}^\dagger a_{\alpha\mathbf{q}}}{2S_\alpha}}\approx \hbar\sqrt{2S_\alpha}a_{\alpha\mathbf{q}}^\dagger\,, \\
S_{\alpha\mathbf{q}}^{z}&= S_\alpha-a_{\alpha\mathbf{q}}^\dagger a_{\alpha\mathbf{q}}\,,
\end{align}
\end{subequations}
where $S_\alpha$ is the total spin in electrode $\alpha$. In the approximation it was assumed that the deviation from spin $S_\alpha$ is small.

Under the transformation intoduced above, magnon Hamiltonian $H_{\rm mag}$ takes the form
\begin{equation}
H_{\rm mag}=\sum_{\alpha\mathbf{q}}\epsilon_{\mathbf{q}}a_{\alpha\mathbf{q}}^{\dagger}a_{\alpha\mathbf{q}},
\end{equation}
where $\epsilon_{\mathbf{q}}$  is the spin wave energy (assumed equal in both reservoirs) for the wavevector $\bf q$, which is given by the formula (see eg. Ref.~\cite{Mattis}) $\epsilon_{\mathbf{q}} =2SJ\sum_{\mathbf{\delta}}[1-\cos(\mathbf{q}\cdot\mathbf{r}_\delta)] + g_m\mu_BB$, with
$\mathbf{r}_\delta$ denoting vectors to nearest neighbors, $S$ and $J$ standing for the localized spin number and exchange constant, and $g_m$ being the Lande factor of the magnetic electrodes.
Note, multimagnon processes are neglected in this approximation.

Finally, the term $H^{\rm t}_{\rm mag}$ can be written as
\begin{equation}
H^{\rm t}_{\rm mag} = \sum_{\alpha\mathbf{q}}j_{\alpha\mathbf{q}}a_{\alpha\mathbf{q}}^{\dagger}d_{\uparrow}^{\dagger}d_{\downarrow}+\mathbf{ \rm{H.c.}},
\end{equation}
where $j_{\alpha\mathbf{q}}$ generally depends on the distribution of interfacial spins and also on coupling between these spins and the quantum dot. The explicit form of $j_{\alpha\mathbf{q}}$ is not required here as this coupling will be treated as a phenomenological  parameter (see below).

\subsection{Method}

In order to calculate magnon leakage current I employ the T-Matrix method to obtain terms up to the second-order in Pauli's Master equation~\cite{Flensberg,Timm},
\begin{align}\label{eq:tmatrix}
T&=T^{(1)}+T^{(2)}+... \nonumber \\
&\approx H_{T}+H_{T}\frac{1}{E_{i}-H_{0}}H_{T}\,,
\end{align}
where $E_{i}$ is the energy of the initial state, $H_T$ denotes the perturbation term ($H^{\rm t}_{\rm mag}$ in our case), and $H_{0}$ is the unperturbed Hamiltonian of the system (Hamiltonian (1) with the term $H^{\rm t}_{\rm mag}$ neglected in our case).

In the sequential tunneling case, i.e., when only term $T^{(1)}$ is taken into account, occupation probabilities $P_i$ of the dot can be derived from the master equation which takes the form
\begin{equation}\label{eq:master}
\dot{P}_{n}=\sum_{m}\left(W^{mn}P_{m}-W^{nm}P_{n}\right)\,,
\end{equation}
where $W^{nm}$ is the transition rate from the dot's state $|n\rangle$ to the state $|m\rangle$. This transition rate is given by the Fermi's golden rule as
\begin{equation}
W^{nm}=\frac{2\pi}{\hbar}\sum_{if}|\langle f|T^{(1)}|i\rangle|^{2}\delta(E_{f}-E_{i})\,,
\end{equation}
where $|i\rangle$ and $|f\rangle$ are the initial and final state of the system, respectively. 
This equation leads to the appearance of magnon populations in the magnonic reservoirs, $\langle a_{\alpha\mathbf{q}}^{\dagger}a_{\alpha\mathbf{q}} \rangle\equiv n_{\alpha}^+(\varepsilon_{\alpha\mathbf{q}})$, which are determined by the Bose-Einstein distribution $n_{\alpha}^+(\varepsilon)=1/\left[\exp\left(\varepsilon/k_BT_\alpha \right) -1\right]=n_\alpha^-(\varepsilon)-1$.

The sequential contribution to total magnon current that flows through the dot  can be calculated from the formula $j_{\rm mag}^{(seq)}=j_{\rm mag,L}^{(seq)}-j_{\rm mag, R}^{(seq)}$, where contribution $j_{\rm mag,\alpha}^{(seq)}$ for transport between electrode $\alpha=L,R$ and the dot is expressed as follows
\begin{equation}
j_{\rm mag,\alpha}^{(seq)} = -\hbar \left(P_{\uparrow}W_\alpha^{\uparrow\downarrow}-P_{\downarrow}W_\alpha^{\downarrow\uparrow}\right)\,.
\end{equation}
In the above expression $W^{\uparrow\downarrow}_\alpha$ and $W^{\downarrow\uparrow}_\alpha$ are the transition rates between states $|\uparrow\rangle$ and $|\downarrow\rangle$ for magnon transfer between electrode $\alpha$ and the dot, while $P_\uparrow$ and $P_\downarrow$ denote probabilities that the quantum dot is in state $|\uparrow\rangle$ and $|\downarrow\rangle$, respectively.

Consider now the second-order processes.
The cotuneling rate can be obtained from Eq.~(\ref{eq:tmatrix}) and the Fermi's golden rule,
\begin{equation}
\Gamma_{\alpha \rightarrow \alpha'}^{mn}=\frac{2\pi}{\hbar}\sum_{if}|\langle f|T^{(2)}|i\rangle|^{2}\delta(E_{f}-E_{i})\,,
\end{equation}
where $\alpha,\alpha'$ denote different electrodes (emitter and absorber of a magnon) and $m,n$ are initial and final states of the dot, with $m=n$ in the case of elastic cotunneling considered in this paper.

Let us consider the magnon cotunneling through a quantum dot occupied with an electron with  spin $\sigma$. The magnon current flowing between the electrodes can be understood as a magnon leakage current, that can play a significant role in such a device in the strong  coupling regime. This process can be understood as follows: when a spin-$\sigma$ electron resides on the dot, the dot can either emit or absorb a magnon from electrode $\alpha$. This process is associated with a flip of the dot's initial spin. However, this spin can flip back to its initial state simultaneously absorbing or emitting a magnon further into the electrode $\alpha'$. The final state of such a transistion can be written down as:
\begin{equation}
|f\rangle = a_{L\mathbf{q}}a_{R\mathbf{q'}}^{\dagger}|i\rangle\,.
\end{equation}

The cotunneling magnon current through the quantum dot can be expressed as,
\begin{equation}
j_{mag,\sigma}^{(cot)}=-\hbar P_{\sigma}\left( \Gamma_{L\rightarrow R}^{\sigma\sigma} - \Gamma_{R\rightarrow L}^{\sigma\sigma} \right)\,,
\end{equation}
where the cotunneling rate is given by the formula:
\begin{equation}
\label{eq:rates}
\Gamma_{L\rightarrow R}^{\sigma\sigma}-\Gamma_{R\rightarrow L}^{\sigma\sigma}=\frac{1}{\hbar}J_{L}J_{R}\int_D \mathop{d\varepsilon} \frac{n_L^{+}(\varepsilon)-n_R^+(\varepsilon)}{\left(\varepsilon-\varepsilon_{\overline{\sigma}\sigma} \right)^2}\,,
\end{equation}
where $\varepsilon_{\overline{\sigma}\sigma}=\varepsilon_{\overline{\sigma}}-\varepsilon_{\sigma}$, $J_{L,R}=2\pi\langle|j_{\mathbf{q}L,R}|\rangle\rho_\alpha$ stand for the effective coupling parameters with $\rho_\alpha$ being density of states in insulating electrodes, and $D$ denotes that the integration is over the magnon band of a finite width.

In contrast to a previous work~\cite{karwackiPRB}, a finite width of the magnon band is assumed, which is more realistic on one side and also is required when considering magnon cotunneling processes. Note, that the spin splitting of the dot level and also exact position of the magnon bottom band edge grow linearly with increasing magnetic field. 
 
In order for sequential transport to occur, magnon energy has to obey inequality $E_{mag}\leq g\mu_BB$, where $E_{mag}=g_m\mu_BB+D_0$. Here, $g_m=\gamma g$ is the $g$-factor of magnonic reservoir assumed proportional to the $g$-factor of quantum dot (with $\gamma$ being the proportionality constant) and $D_0$ is the bottom edge of the magnon band for $B=0$ (e.g. due to a magnetic anisotropy). From this follows, that there is a critical magnetic field $B_c$ for sequential processes to occur,
\begin{equation}
g\mu_BB_c=\frac{D_0}{1-\gamma}\,,
\end{equation}
Note that when both g-factors are equal, i.e. $\gamma=1$, no sequential transport can occur. More generally, no sequential transport can occur for $B<B_c$.

When calculating the cotunneling contribution to magnon current, one encounters a singularity at $\varepsilon=\varepsilon_{\overline{\sigma}\sigma}$ for $B>B_c$, so one has to calculate the so-called Hadamard ($\mathcal{H}$) finite part of the appropriate integral, defined as:
\begin{equation}
\mathcal{H}\int_D \mathop{d\varepsilon} \frac{n_L^+(\varepsilon)-n_R^{+}(\varepsilon)}{\left(\varepsilon-\varepsilon_{\overline{\sigma}\sigma} \right)^2}=\frac{\operatorname{d}}{\operatorname{d\varepsilon_{\overline{\sigma}\sigma}}}\mathcal{C}\int_D \mathop{d\varepsilon} \frac{n_L^+(\varepsilon)-n_R^{+}(\varepsilon)}{\varepsilon-\varepsilon_{\overline{\sigma}\sigma} }\,,
\end{equation}
where $\mathcal{C}$ denotes the Cauchy's principal part.

\section{Numerical results}

In this section numerical results on magnon transport in the system under consideration are shown. The focus is on the influence of magnetic field $B$ and  difference $\Delta T$ in temperatures of the two magnonic reservoirs on the sequential and cotunneling magnon currents. Especially interesting is the case of magnetic field below the critical  field $B_c$, when only cotunneling processes can occur. In the following I assume $\gamma=0.9$ and the magnon band edge (due to anisotropy) $D_0=k_BT_0$.

\subsection{Case of both sequential and cotunneling currents ($B>B_c$)}

\begin{figure}[!t]
\centering
\includegraphics[width=\columnwidth]{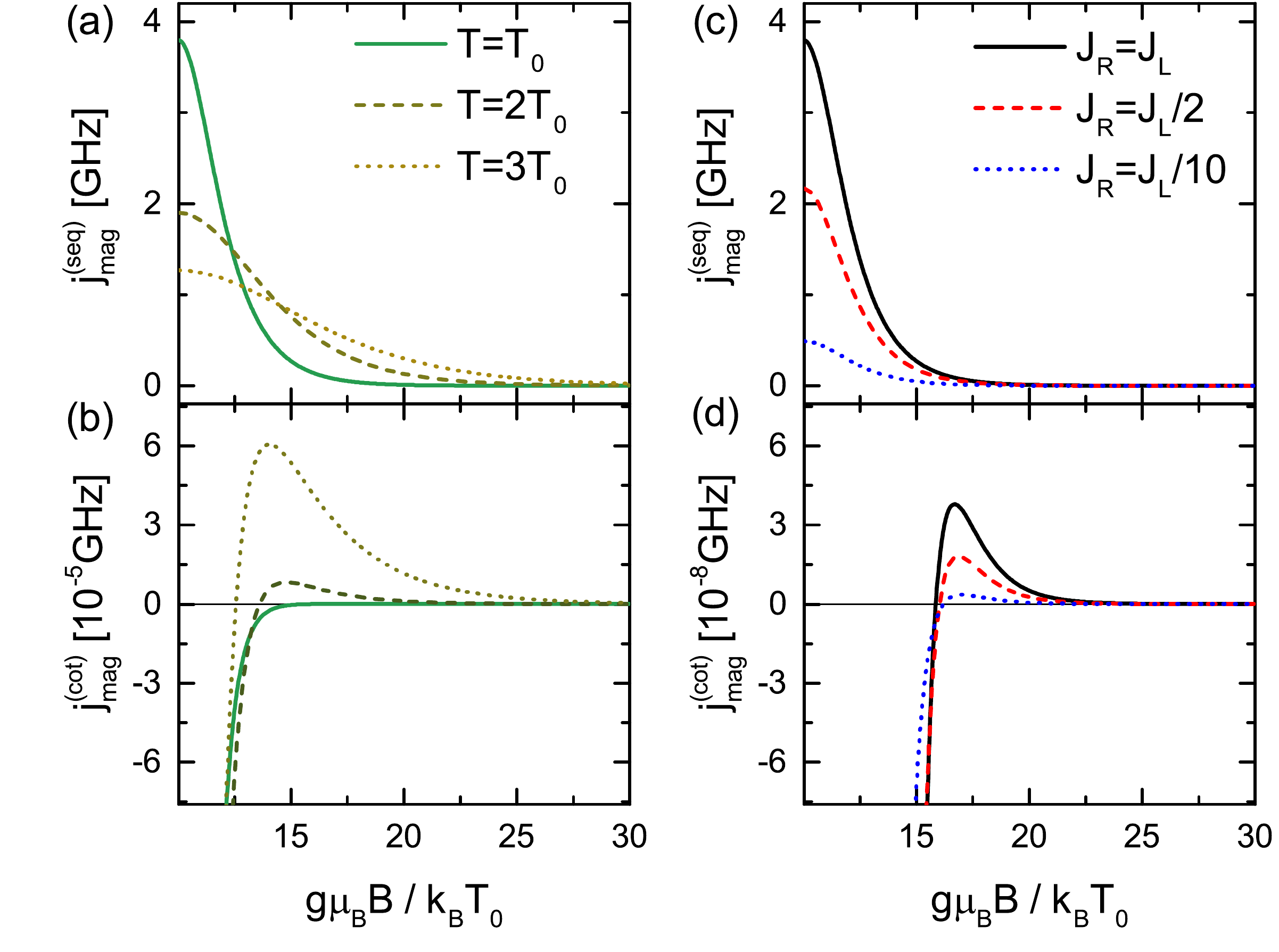}
\caption{Sequential contribution to magnon current, $j_{mag}^{(seq)}$, and cotunneling contribution to magnon current, $j_{mag}^{(cot)}$, as a function of magnetic field $g\mu_BB$ for indicated values of the the average temperature $T$ of magnon reservoirs [(a) and (b)], and for indicated values of the coupling parameter $J_R$ [(c) and (d)]. Other parameters (unless otherwise specified in figure): $\Delta T=T_0$, $J_L=J_R=0.1k_BT_0$, $T=T_0$, $\gamma=0.9$, $D_0=k_BT_0$, and $k_BT_0=0.1$~meV.}
\label{fig:fig2}
\end{figure}

Firstly, the case when both sequential and cotunneling processes contribute to the magnon current is considered. This means that the magnetic field $B$ is larger than the critical magnetic field $B_c$ corresponding to the assumed values of $\gamma$ and $D_0$.
It is assumed that the dot is initially prepared in one
of the two spin states, $\lvert\uparrow\rangle$ or $\lvert\downarrow\rangle$. Such a single-electron state is required
in order to mediate magnon transport between the two magnonic reservoirs.

Transport of magnons through the dot does not change its charge state, so the dot remains singly occupied
and only its  spin state can vary due to magnon sequential processes. The corresponding stationary occupation
probabilities, $P_{\uparrow}$ and $P_{\downarrow}$, can be found from equation~(\ref{eq:master}) which can be expressed in matrix form,
\begin{equation}
\widetilde{\mathbf{W}}\mathbf{P}=\mathbf{0}\,,
\end{equation}
with $\mathbf{0}$ being vector of zeros. Additionally, a probability conservation, $P_\uparrow+P_\downarrow=1$, is taken into account.
Since the cotunneling processes are elastic  and the corresponding changes in the spin state of the dot are only virtual, the probabilities depend only on sequential tunneling rates.
To solve this equation matrix $\widetilde{\mathbf{W}}$ is necessary, which in the case under consideration acquires the form
\begin{equation}
\widetilde{\mathbf{W}}=\frac{1}{\hbar}\sum_{\alpha}
\left[ \begin{array}{cc}
-J_{\alpha}n_{\alpha}^{+} & J_{\alpha}n_{\alpha}^{-} \\	
J_{\alpha}n_{\alpha}^{+} & -J_{\alpha}n_{\alpha}^{-}
\end{array}\right]\xi(|g\mu_BB|)\,,
\end{equation}
where
\begin{equation}
\xi(|g\mu_BB|)=\begin{cases}
0 & g\mu_BB < g\mu_BB_c    \\
1 & g\mu_BB \geq g\mu_BB_c
\end{cases}\,.
\end{equation}
The above function $\xi(|g\mu_BB|)$ introduces dependence of the sequential processes on the critical magnetic field derived in previous section.

In Fig.~\ref{fig:fig2}(a) and (b) the sequential and cotunneling currents as a function of external magnetic field $B$ for indicated values of the average temperature  $T$ of magnetic reservoirs are shown. Since sequential magnon transport has been described in more detail elsewhere~\cite{karwackiPRB}, only the most important aspects of sequential magnon transport are highlighted and focus is rather on the cotunneling contribution to transport.

Magnon current flows from the electrode with higher temperature to the one with lower temperature. An individual sequential process changes state of the dot from $|\sigma\rangle$ to $|\overline{\sigma}\rangle$. The maximum of sequential current corresponds to low-energetic magnons, i.e., $\varepsilon=g\mu_BB\approx g\mu_BB_c$, where according to Bose-Einstein distribution the average number of magnons is the largest. With the increase in Zeeman splitting of the quantum dot's energy level, only highly energetic magnons contribute to transport. When average temperature $T$ of the magnonic reservoirs increases (with the temperature difference $\Delta T$ of the reservoirs kept constant), the maximal magnon current (corresponding to the low-energetic magnons) decreases.

According to Fig.~\ref{fig:fig2}(b), the cotunneling current is few orders of magnitude smaller, and also behaves differently with magnetic field. Note, Fig.~\ref{fig:fig2} corresponds to $B>B_c$. If $g\mu_BB> 10k_BT_0$ the cotunneling current increases until it reaches a maximum. In contrast to the sequential magnon  current, maximal value of the cotunneling current increases with the increasing average temperature of the reservoirs. This results from the fact, that higher temperature leads to a larger number of  spin wave excitations in the electrodes and, thus, the difference $n_L^+(\varepsilon)-n_R^+(\varepsilon)$ also increases. Note, the total magnon current is positive, though in some range of magnetic field the cotunneling contribution is negative. 
Increasing coupling asymmetry, on the other hand, results in a decrease in both sequential and cotunneling currents, as shown in Fig.~\ref{fig:fig2}(c) and (d).

\subsection{Case of suppressed sequential transport ($B<B_c$)}

\begin{figure}[!t]
\centering
\includegraphics[width=\columnwidth]{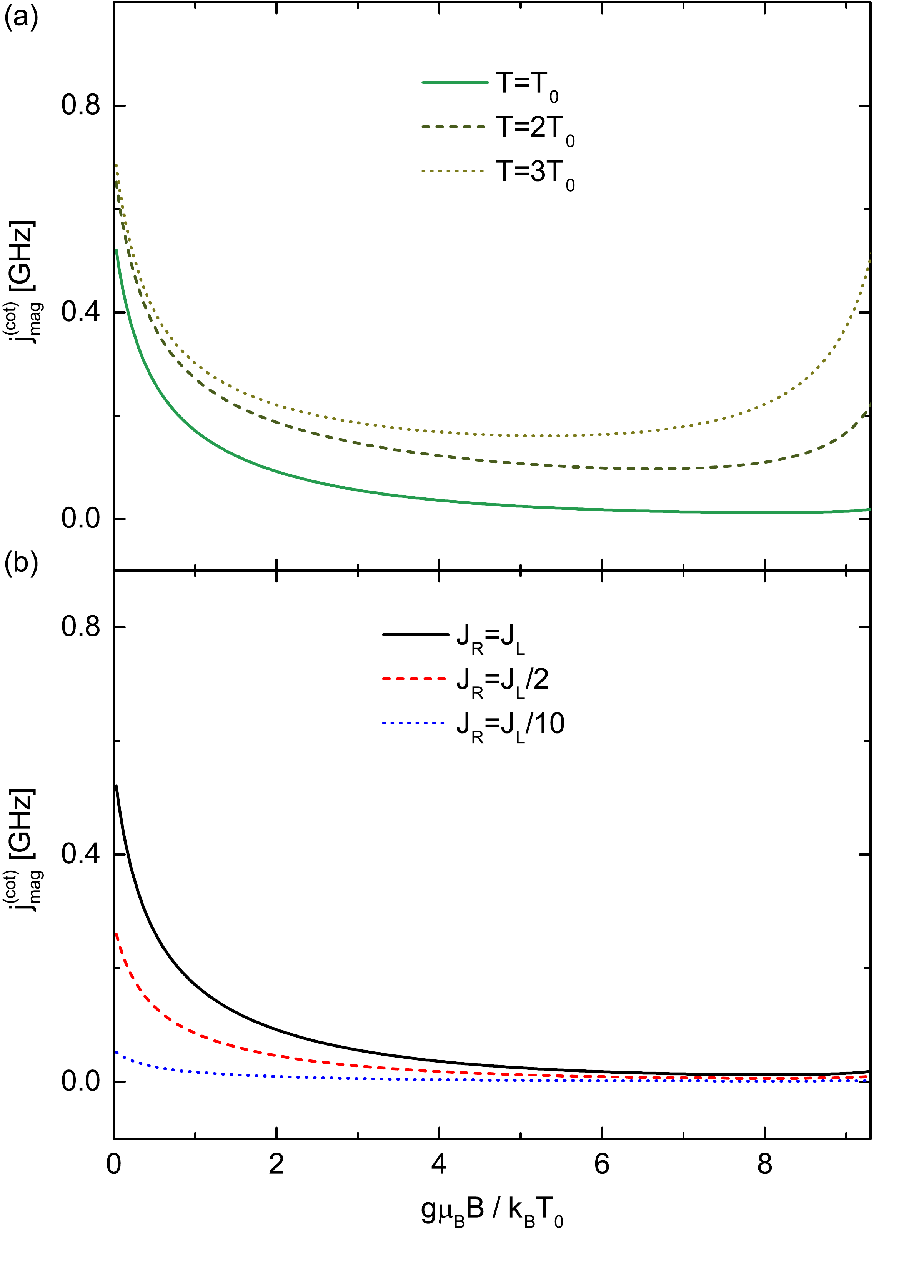}
\caption{Cotunneling contribution to magnon current, $j_{mag}^{(cot)}$, as a function of magnetic field $g\mu_BB$ for indicated values of the average temperature $T$ of reservoirs and of the coupling parameter $J_R$. Other parameters (unless otherwise specified in figure): $J_L=J_R=0.1k_BT_0$, $T=T_0$, $\gamma=0.9$, $D_0=k_BT_0$, and $k_BT_0=0.1$~meV.}
\label{fig:fig3}
\end{figure}

Now consider the situation when magnetic field is below the critical value, which means that the energy matching for sequential transport cannot be obeyed. However, a small leakage current can flow through the system due to magnon cotunneling. It is assumed that the $\lvert\uparrow\rangle$ and $\lvert\downarrow\rangle$ spin states of the dot are equally probable, and these probabilities  are constant due to absence of  sequential transport.

In this regime, the cotunneling magnon current, shown as a function of magnetic field in Fig.~\ref{fig:fig3}, changes rapidly for small magnetic fields, i.e. $B\approx 0$, and when the field $B$ approaches  the critical magnetic field, $B=B_c$. This follows from the Bose-Einstein distribution function. When the average temperature of both reservoirs increases, see  Fig.~\ref{fig:fig3}(a), there is an increase in cotunneling magnon current, similarly as in the case described in the previous section, i.e. higher temperature leads to a larger difference $n_L-n_R$.

Increasing asymmetry in the coupling between the dot and the two electrodes results in a monotonic decrease of the current, as shown in Fig.~\ref{fig:fig3}(b). When one  of the reservoirs is completely decoupled from the dot, i.e. $J_R\rightarrow 0$, no cotunneling current can flow in the system.

\begin{figure}[!t]
\centering
\includegraphics[width=\columnwidth]{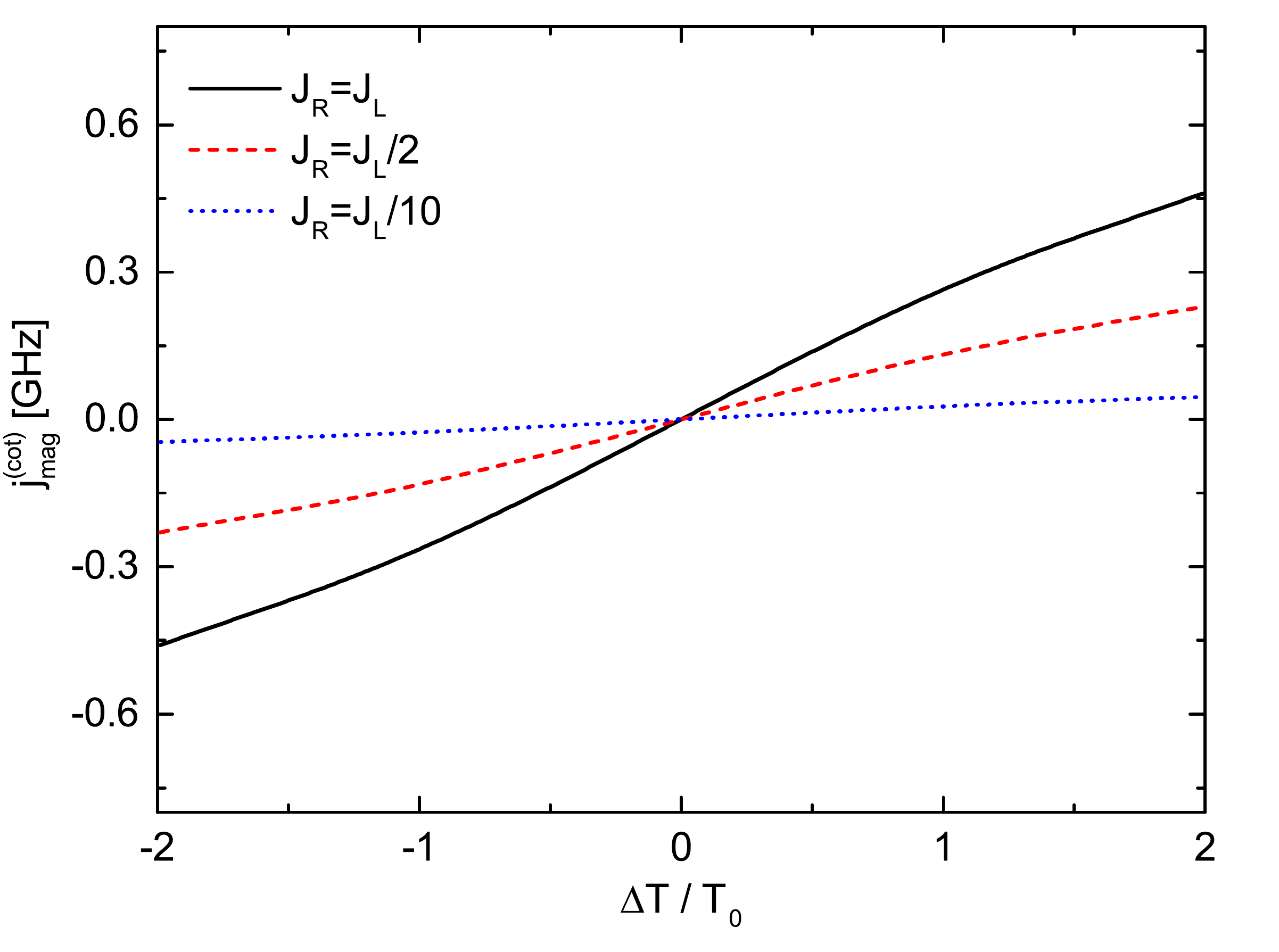}
\caption{Cotunneling contribution to magnon current, $j_{mag}^{(cot)}$, as a function of reservoir temperature difference $\Delta T$ for indicated values of coupling parameter $J_R$. Other parameters (unless otherwise specified in figure): $T=T_0$, $g\mu_BB=k_BT_0/2$, $\gamma=0.9$, $D_0=k_BT_0$, and $k_BT_0=0.1$~meV.}
\label{fig:fig4}
\end{figure}

Fig.~\ref{fig:fig4} shows the cotunneling current as a function of temperature difference between magnonic reservoirs for $B<B_c$. When both reservoirs have the same temperature, average magnon current is zero and leakage magnon (spin) current through the system appears for  $|\Delta T| >0$. The cotunneling current is maximal for both electrodes equally coupled to the dot and it decreases monotonically with the increase in coupling asymmetry. For positive $\Delta T$ the magnons leak from left electrode into the right one, while the opposite is true for the case when right electrode is of higher temperature than the left one. In contrast to sequential transport there can be no diode effect in cotunneling regime due to coupling asymmetry.

\section{Summary}
In conclusion, I have shown that in quantum dot systems coupled to magnetic insulators not only sequential magnon transport is possible, but a small cotunneling contribution can play a role as well. This contribution is dominant in the case with some mismatch between the quantum dots's and insulator's g-factors, when only a cotunneling current may flow. Such a cotunneling magnon  current can be then understood as a spin leakage current. This leakage current is highly sensitive to changes in such parameters of the system as temperature and the coupling asymmetry. Further investigation may focus on the influence of such a spin leakage current on the conversion between spin currents of magnonic and electronic nature.

\section*{Acknowledgement}
This work was supported by National Science Centre in Poland as Project No. DEC-2012/04/A/ST3/00372. The author would like to thank J{\'o}zef Barna{\'s} for providing valuable comments that greatly improved the manuscript.

\section*{References}

\bibliography{mybibfile}

\end{document}